# Effect of the nature of alkali and alkaline-earth oxides on the structure and crystallization of an aluminoborosilicate glass developed to immobilize highly concentrated nuclear waste solutions

A. Quintas[1], D. Caurant[1], O. Majérus[1], T. Charpentier[2], J-L. Dussossoy[3]

[1] *Laboratoire de Chimie de la Matière Condensée de Paris (UMR 7574),*
*Ecole Nationale Supérieure de Chimie de Paris (ENSCP, ParisTech), Paris, 75005, France*

[2] *CEA Saclay, Laboratoire de Structure et Dynamique par Résonance Magnétique,*
*DSM/DRECAM/SCM – CEA CNRS URA 331, Gif-sur-Yvette, 91191, France*

[3] *Laboratoire d'Etude de Base sur les Verres, CEA Valrho, DEN/DTCD/SCDV/LEBV,*
*Bagnols-sur-Cèze, 30207, France*

*Corresponding author: daniel-caurant@enscp.fr*

**Abstract** *A complex rare-earth rich aluminoborosilicate glass has been proved to be a good candidate for the immobilization of new high level radioactive wastes. A simplified seven-oxides composition of this glass was selected for this study. In this system, sodium and calcium cations were supposed in other works to simulate respectively all the other alkali ($R^+=Li^+, Rb^+, Cs^+$) and alkaline-earth ($R'^{2+}=Sr^{2+}, Ba^{2+}$) cations present in the complex glass composition. Moreover, neodymium or lanthanum are used here to simulate all the rare-earths and actinides occurring in waste solutions. In order to study the impact of the nature of $R^+$ and $R'^{2+}$ cations on both glass structure and melt crystallization tendency during cooling, two glass series were prepared by replacing either $Na^+$ or $Ca^{2+}$ cations in the simplified glass by respectively ($Li^+, K^+, Rb^+, Cs^+$) or ($Mg^{2+}, Sr^{2+}, Ba^{2+}$) cations. From these substitutions, it was established that alkali ions are preferentially involved in the charge compensation of $(AlO_4)^-$ entities in the glass network comparatively to alkaline-earth ions. The glass compositions containing calcium give way to the crystallization of an apatite silicate phase bearing calcium and rare-earth ions. The melt crystallization tendency during cooling strongly varies with the nature of the alkaline-earth.*

## INTRODUCTION

New confinement glasses, aimed at immobilizing highly concentrated nuclear waste solutions, stemming from the reprocessing of high discharge burn up nuclear fuel (60 GWj.t$^{-1}$), are currently under study. For instance, a complex rare-earth rich aluminoborosilicate glass (16 wt% rare-earth oxides) has already proved from a technological point of view (chemical durability, waste capacity and low crystallization tendency), to be a good candidate for the immobilization of these concentrated high level radioactive wastes [1,2]. High amounts of alkali ($R^+$) and alkaline-earth ($R'^{2+}$) cations are present in this glass composition coming either from the waste or from the glass frit mixed with the waste and are incorporated in glass structure. For instance, according to the simulation results of high discharge burn up spent fuel composition [3], $Rb_2O$, $Cs_2O$, SrO and BaO ions will represent about 17 wt% of all the fission products in waste solutions. A high quantity of sodium will also occur in waste solutions and stems from nuclear spent fuel reprocessing. Both sodium and lithium will be also added in glass frit composition to facilitate nuclear glass fabrication and waste incorporation. To increase the chemical durability of borosilicate nuclear glasses, calcium is also introduced in their composition. In the complex rare earth-rich aluminoborosilicate glass considered here, $R_2O$ and R'O concentrations (in mol.%) are the following [2]: 8.49 ($Na_2O$), 5.36 (CaO), 5.03 ($Li_2O$), 0.44 (BaO), 0.33 ($Cs_2O$), 0.30 (SrO), 0.07 ($Rb_2O$). In previous works [1,4,5], we mainly studied the structure and crystallization tendency of a simplified seven-oxide rare-earth rich aluminoborosilicate glass ($SiO_2$, $B_2O_3$, $Al_2O_3$, R'O, $R_2O$, $ZrO_2$, $RE_2O_3$ with RE=La or Nd) derived from the complex glass composition by using the most abundant $R'^{2+} = Ca^{2+}$ and $R^+ = Na^+$ cations to simulate all the alkaline-earth and alkali cations occurring in wastes. Nevertheless several results concerning the crystallization tendency of the complex glass with all alkali and alkaline-earth cations were given in [2,3]. In this paper, we present the impact of totally changing the nature of alkali ($R^+ = Li^+, Na^+, K^+, Rb^+, Cs^+$) or alkaline-earth ($R'^{2+} = Mg^{2+}, Ca^{2+}, Sr^{2+}, Ba^{2+}$) cations on both the structure and the crystallization tendency of the simplified glass. Even if $K^+$ and $Mg^{2+}$ cations will be not present in the complex nuclear glass composition, their effect on glass structure was also studied in this work to complete the alkaline-earth and alkali cations series. The effect of the nature of $R^+$ and $R'^{2+}$ cations and of their relative proportions on the environment of $RE^{3+}$ cations was presented in [6] where we showed that both alkali and alkaline earth cations are present around $Nd^{3+}$ ions in non-bridging oxygen atoms (NBOs)-rich regions enabling their stabilization in glass structure.

**EXPERIMENTAL PROCEDURE**

The simplified seven-oxides glass composition studied in this work is: 61.79 $SiO_2$ – 3.05 $Al_2O_3$ - 8.94 $B_2O_3$ – 14.41 $R_2O$ – 6.32 R'O – 1.89 $ZrO_2$ – 3.60 $RE_2O_3$ (mol.%) (glass A) with RE=La or Nd. Two glass series were prepared for this paper. For the first series (R series), the nature of $R^+$ cation was varied from $Li^+$ to $Cs^+$ (R'= Ca). For the second series (R' series), the nature of $R'^{2+}$ cation was varied from $Mg^{2+}$ to $Ba^{2+}$ (R=Na). All glasses were melted between 1300 and 1400°C in Pt crucibles, quenched and annealed near $T_g$. The structure of all glasses with RE = La was studied by $^{11}B$ (160.14 MHz), $^{27}Al$ (130.06 MHz) and $^{23}Na$ (132.03 MHz) MAS NMR. The chemical shifts ($\delta_{iso}$) of the different nuclei are reported in ppm relative to an external sample of 1 M aqueous boric acid at 19.6 ppm, 1 M aqueous NaCl at 0 ppm and 1 M aqueous $Al(NO_3)_3$ at 0 ppm. In order to simulate the natural cooling of nuclear glasses, all samples were slowly cooled (1°C/min) from 1350°C to room temperature and were then studied by X-ray diffraction (XRD), scanning electron microscopy (SEM) and electron probe microanalysis (EPMA).

**STRUCTURAL STUDY**

Concerning the silicate network, Raman results (not presented in this paper, [7]) show that increasing $R^+$ or $R'^{2+}$ cations field strength induce a displacement to the right of the $2Q_3 \leftrightarrow Q_2 + Q_4$ equilibrium in the melt ($Q_n$ represents $SiO_4$ units with 4-n NBOs) in accordance with the increasing glass-in-glass phase separation tendency with cation field strength in binary silicate glasses. According to [8], the field strengths of alkali ($Li^+, Na^+, K^+, Rb^+, Cs^+$) and alkaline-earth ($Mg^{2+}, Ca^{2+}, Sr^{2+}, Ba^{2+}$) cations in silicate glasses are respectively (0.26, 0.18, 0.12, 0.11, 0.10) and (0.46, 0.36, 0.29, 0.26).

No other environment than fourfold coordinated aluminum units was detected by MAS and MQMAS $^{27}Al$ NMR (data not shown) in all glasses. The absence of significant evolution between $^{27}Al$ MAS NMR spectra (the corresponding NMR parameters are given Fig.1) when the nature of alkaline-earth is varied (and with R=Na) indicates that $(AlO_4)^-$ units are preferentially charge compensated by $Na^+$ cations rather than by $R'^{2+}$ cations. Indeed, $^{27}Al$ quadrupolar coupling constant $C_Q$ is known to increase with the field strength of charge compensator [9]. Comparison of $^{27}Al$ MAS NMR results of glasses of the R' series with that of a glass derived from glass A (referred to as R0 in Fig.1) but containing only $Na^+$ ions confirms that for all glasses of the R' series, the great majority of four-coordinated aluminum are charge compensated by $Na^+$ ions within the limit of MAS NMR sensitivity. On the other hand, significant evolution of both $C_Q$ and $\delta_{iso}$ are observed between the samples of the R series when $R^+$ cation field strength increases (Fig.1) showing that the nature of the charge compensators of $(AlO_4)^-$ units strongly changes for this series.

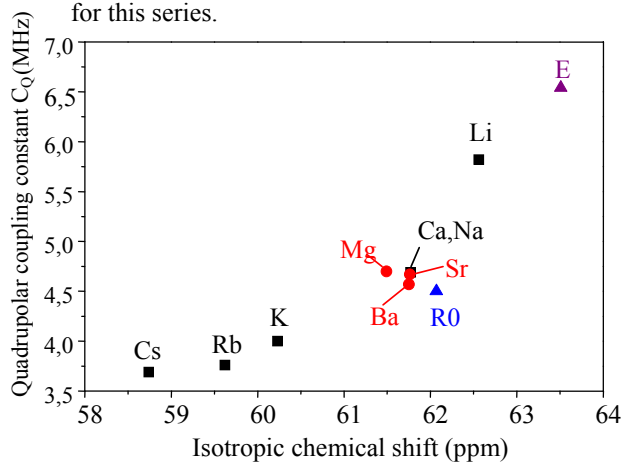

**Fig. 1.** Evolution of the mean $C_Q$ and $\delta_{iso}$ parameters deduced from $^{27}Al$ MAS NMR spectra for R and R' series. For comparison, the parameters of a glass (R0) derived from glass A but containing only $Na^+$ ions and of an industrial calcium aluminoborosilicate glass (E) for which $AlO_4^-$ units are essentially compensated by $Ca^{2+}$ ions are also shown. MAS NMR data were analyzed according to the method described in [5].

As for the samples containing $K^+$, $Rb^+$ or $Cs^+$ ions both $C_Q$ and $\delta_{iso}$ decrease with $R^+$ cation field strength in comparison with sodium, this shows that for these three glasses the great majority of $(AlO_4)^-$ units remain charge compensated by alkali cations. Indeed, the presence of high field strength $Ca^{2+}$ ions as aluminum charge compensator should have induced significant increase of $^{27}Al$ NMR parameters (see Fig.1 the NMR parameters of the industrial calcium aluminoborosilicate glass (E-glass) for which $AlO_4^-$ units are essentially compensated by $Ca^{2+}$ ions). Concerning the glass with $Li^+$, the replacement of all $Na^+$ by $Li^+$ ions induces an important increase of both $^{27}Al$ NMR parameters (Fig.1). The question is: are $(AlO_4)^-$ units still preferentially charge compensated by alkali ions (here $Li^+$) in this glass or does $Ca^{2+}$ ions participate to charge compensation? The fact that for R=Li, $C_Q$ and $\delta_{iso}$ parameters remain lower than E-glass ones shows that lithium participate to aluminum charge compensation. Moreover, the comparison of the quadrupolar coupling constant $C_Q$ of the glass with Li ($C_Q$=5.8 MHz) with that of a lithium aluminoborosilicate glass with $Li^+$ ions as unique charge compensators of $(AlO_4)^-$ units ($C_Q$=6.05 MHz) [10] seems to indicate that the aluminum tetrahedral units are preferentially charge compensated by $Li^+$ ions in our glass.

Concerning $^{11}B$ MAS NMR results, the evolution of spectra and of the proportion $N_4$ of $(BO_4)^-$ units for the R' series is shown in Fig. 2.

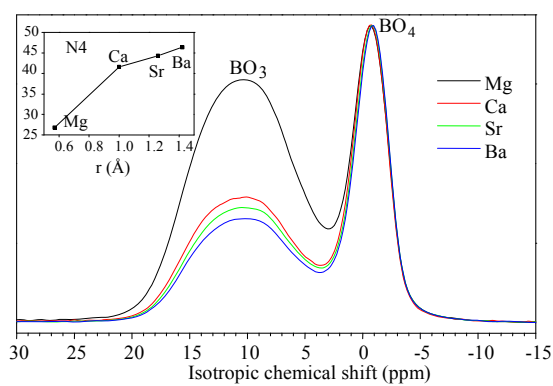

**Fig. 2.** $^{11}B$ MAS NMR spectra of glasses of the R' series. The evolution of the proportion $N_4$ of $(BO_4)^-$ units versus cation radius is also shown. All spectra are normalized on the maximum of the $BO_4$ peak.

It appears that the substitution of an alkaline earth by a one with a lower field strength leads to an increase of $N_4$. Thus, the field strength of $R'^{2+}$ cations seems to control $N_4$ evolution for this series. Similar evolution was observed by replacing calcium by sodium in glass A [5]. Moreover, the small $\delta_{iso}$ variation of $(BO_4)^-$ units between -0.66 and -0.53 ppm for the R' series remains close to that of $(BO_4)^-$ units charge compensated only by $Na^+$ ions as in glass A with only sodium ($\delta_{iso}$=-0.64 ppm) and far from that of $(BO_4)^-$ units charge compensated only by $Ca^{2+}$ ions as in industrial E-glass ($\delta_{iso}$=-0.11 ppm). This shows that tetrahedral boron units remain preferentially charge compensated by $Na^+$ ions for all glasses of R' series. However, for the R series, the evolution of $N_4$ is not monotonous when $R^+$ cation field strength decreases and the reason for this seems to be complex.

Thus, all these structural results show that the environment of $(AlO_4)^-$ units will be probably not affected by replacing strontium and barium by calcium in the complex nuclear glass. Moreover, due to the small SrO and BaO concentrations in complex glass composition (see above), $N_4$ will probably increase only very slightly by introducing these oxides in the melt. In the complex nuclear glass, $(AlO_4)^-$ units will be mainly compensated by $Na^+$ ions because $[Na_2O]$=8.49 > $[Al_2O_3]$=3.05 >>$[Rb_2O]$+$[Cs_2O]$ =0.40 mol%. Nevertheless, because of the significant $Li_2O$ concentration ($[Li_2O]$=5.03 mol%) in complex glass, lithium will probably also participate to the charge compensation of tetrahedral aluminum units. According to $^{11}B$ NMR results [5,7], the $(BO_4)^-$ units charge compensation will be probably preferentially insured by $Na^+$ and $Li^+$ ions. Nevertheless, $Ca^{2+}$ ions will also certainly contribute to this charge compensation.

**CRYSTALLIZATION STUDY**

Our previous studies [4] performed on glass A with RE=La or Nd and containing only CaO and $Na_2O$ with various K=$[CaO]$/($[Na_2O]$+$[CaO]$) ratios showed that the crystallization of a silicate apatite phase with $Ca_{2+x}RE_{8-x}(SiO_4)_6O_{2-0.5x}$ composition (x ~ 0.4 to 0.7) occurred (mainly from sample surface) during slow cooling. The crystallization tendency increased with K and by replacing La by Nd. Moreover, sodium was not detected in apatite crystals.

The XRD patterns and pictures of the R' series (RE=Nd) samples recovered after slow cooling (1°C/min) are shown in Fig.3. Apatite crystallization was only detected when R'=Ca. When R'=Sr or Ba no crystallization is detected (this was confirmed by SEM) whereas an unidentified crystalline phase is detected for the sample containing Mg.

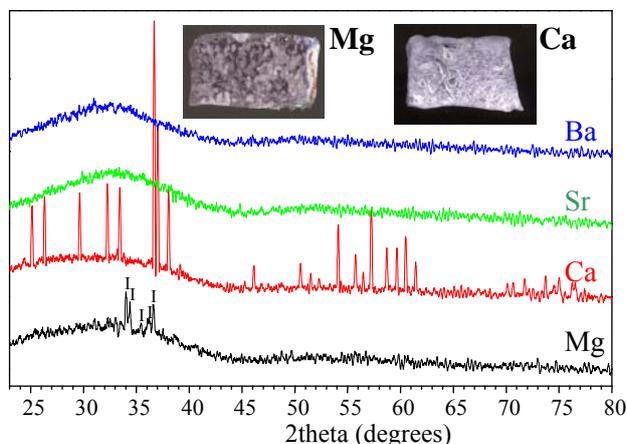

Fig.3. XRD patterns of slowly cooled samples of the R' series with RE=Nd ($\lambda_{K\alpha1}$Co=0.1789 nm). For R'=Ca all peaks are attributed to apatite whereas for R'=Mg an unidentified crystalline phase (I) is detected. Pictures of samples for R'=Mg and Ca are shown in the upper part.

Consequently, the nature of $R'^{2+}$ cation has a very strong effect on the crystallization tendency of the undercooled melt. The introduction of small SrO and BaO amounts in the complex glass composition is thus not expected to induce the crystallization of new phases. This is in agreement with our results reported in [2] concerning the crystallization of the complex nuclear glass (with both CaO, SrO, MgO, Na$_2$O, Li$_2$O, Cs$_2$O) for which only the calcium apatite phase was observed after slow cooling at 1°C/min.

For the R series samples (that all contain CaO) with RE=Nd, XRD, SEM and EPMA shows that the calcium apatite Ca$_{2+x}$Nd$_{8-x}$(SiO$_4$)$_6$O$_{2-0.5x}$ (x ~ 0.14 to 0.41) phase crystallizes in all the samples during slow cooling. Moreover, for R=Li and Cs, other crystalline phases are detected by XRD and SEM, respectively cristobalite (SiO$_2$) and pollucite (CsAlSi$_2$O$_6$) (Fig.4.). The crystallization of pollucite is directly linked to the presence of 14.41 mol% of Cs$_2$O in the R=Cs glass. Because of the low Cs$_2$O concentration (0.33 mol%), pollucite does not crystallize in the complex glass [2]. The dark droplets observed in Fig. 4a (R=Li) probably contains the cristobalite crystals and would originate from glass-in-glass phase separation due the presence of a high concentration of Li$^+$ cations. Glass-in-glass phase separation and cristobalite crystallization were not observed in the complex glass [2] probably because of the lower Li$_2$O concentration in this glass.

The replacement of Na by K or Rb has only small effect on the tendency of the undercooled melt to crystallize (apatite only) because R$^+$ cations are not incorporated into apatite crystals.

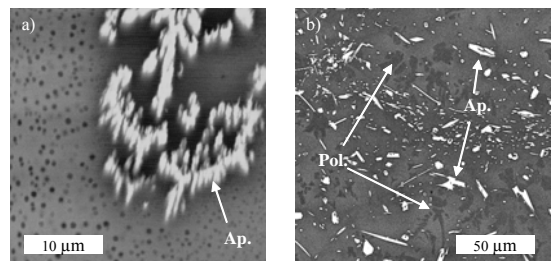

Fig.4. SEM images (backscattered electrons) of slowly cooled samples of the R series: (a) R= Li, (b) R=Cs. Ap: apatite, Pol: pollucite